\documentclass[aps,prb,twocolumn,superscriptaddress,amsmath,amssymb]{revtex4-2}

\usepackage{amssymb}
\usepackage{graphicx}
\usepackage{tabularx}
\usepackage{bm}
\usepackage{epsfig}
\usepackage[ansinew]{inputenc}

\begin{document}

\title{Near-room-temperature ferrimagnetism and half-metallicity in disordered Ca$_{1.5}$La$_{0.5}$MnRuO$_6$}

\author{A. G. Silva}
\affiliation{Instituto de F\'{\i}sica, Universidade Federal de Goi\'{a}s, 74001-970, Goi\^{a}nia, GO, Brazil}

\author{R. B. Pontes}
\affiliation{Instituto de F\'{\i}sica, Universidade Federal de Goi\'{a}s, 74001-970, Goi\^{a}nia, GO, Brazil}

\author{M. Boldrin}
\affiliation{Instituto de F\'{\i}sica, Universidade Federal de Goi\'{a}s, 74001-970, Goi\^{a}nia, GO, Brazil}

\author{H. V. S. Pessoni}
\affiliation{Instituto de F\'{\i}sica, Universidade Federal de Goi\'{a}s, 74001-970, Goi\^{a}nia, GO, Brazil}

\author{L. S. I. Veiga}
\affiliation{Diamond Light Source, Chilton, Didcot, Oxfordshire, OX11 0DE, United Kingdom}

\author{J. R. Jesus}
\affiliation{Centro Brasileiro de Pesquisas F\'{\i}sicas, 22290-180, Rio de Janeiro, RJ, Brazil}

\author{H. Fabrelli}
\affiliation{Centro Brasileiro de Pesquisas F\'{\i}sicas, 22290-180, Rio de Janeiro, RJ, Brazil}

\author{A. R. C. Gonzaga}
\affiliation{Centro Brasileiro de Pesquisas F\'{\i}sicas, 22290-180, Rio de Janeiro, RJ, Brazil}

\author{E. M. Bittar}
\affiliation{Centro Brasileiro de Pesquisas F\'{\i}sicas, 22290-180, Rio de Janeiro, RJ, Brazil}

\author{L. Bufai\c{c}al}
\email{lbufaical@ufg.br}
\affiliation{Instituto de F\'{\i}sica, Universidade Federal de Goi\'{a}s, 74001-970, Goi\^{a}nia, GO, Brazil}

\date{\today}

\begin{abstract}

The electron spin polarization on half-metallic double-perovskites is usually conditioned to the ordered rock-salt arrangement of the transition-metal ions along the lattice. In this work, we investigate a polycrystalline sample of the Ca$_{1.5}$La$_{0.5}$MnRuO$_6$ compound by employing x-ray powder diffraction, high resolution transmission electron microscopy, x-ray absorption and magnetic circular dichroism at the Mn-$L_{2,3}$ and Ru-$M_{2,3}$ edges, magnetometry, electrical transport and first principles calculations in order to show that this is a fully disordered material exhibiting near room temperature ferrimagnetism and half-metallic conductivity, with significant intergrain tunneling magnetoresistance. These unprecedented results are compared to those of archetypical ordered double-perovskites, and discussed in terms of the Mn and Ru valences and their orbital hybridization.

\end{abstract}


\maketitle

\section{Introduction}

Oxides with double-perovskite (DP) structure (A$_2$BB'O$_6$, with A = alkaline-earth/rare-earth, B,B' = transition-metal - TM) are known since the 1950s, however a greater interest in this family of compounds was sparked after the discovery of room-temperature ferrimagnetic (FIM) and half-metallic (HM) behavior in Sr$_2$FeMoO$_6$ \cite{Tokura} and Sr$_2$FeReO$_6$ \cite{Kobayashi} at the end of the 1990s. Later on, several other HM DPs were produced, such as Sr$_2$CrReO$_6$ and Sr$_2$CrOsO$_6$, some of them presenting even higher ferromagnetic (FM)-type transition temperature ($T_C$) than that found for the FeMo- and FeRe-based systems \cite{Serrate,Dasgupta,Tang}. The great majority of HM DPs results from a combination between localized 3$d$ and the relatively delocalized 4$d$/5$d$ TM ions, where mixed valence states in the TM ions are commonly observed on the most prominent compounds \cite{Sami,Serrate}. 

A common problem found on the high temperature DPs prepared by conventional methods of synthesis is the presence of some amount of anti-site disorder (ASD) at the B/B' sites. This means that some of the TM ions switch places along the lattice. In the case of the FM-type HM DPs, this unavoidably disrupts the long-range magnetic order and suppress the 100\% spin-polarization of the charge carriers. This is because the 3$d$--O--3$d$ interactions are typically strongly antiferromagnetic (AFM) and result in non-polarized itinerant electrons in the otherwise insulating 3$d$ spin-channels. This represents a hindrance for  technological applications, since the production of fully ordered DPs requires expensive and time-consuming methods of synthesis. 

An exception to this trend of ASD-induced degradation of half-metallicity is the A$_2$MnRuO$_6$ system, for which long-range magnetic order could be maintained even with severe ASD because, besides the FIM ascribed to the Mn--O--Ru interaction, the ASD-induced Mn--O--Mn coupling can be FM, depending on the Mn oxidation state. Indeed, first principles calculations performed for a completely disordered Ca$_2$MnRuO$_6$ compound predict FIM ordering due to AFM-favored Mn--O--Ru coupling, while the HM conductivity is ensured by the fact that the Mn--O--Mn interaction brought by ASD is FM-favored due to Mn$^{3+}$/Mn$^{4+}$ mixed valence \cite{Mishra}. However, the experimental observations on this disordered orthorhombic crystal indicate a tendency toward Mn$^{4+}$/Ru$^{4+}$ configuration \cite{Zhou,Ricciardo,King}. This means it lacks clear mixed valence character, which is desirable for two reasons: (i) the double-exchange FM coupling on the Mn$^{3+}$--O--Mn$^{4+}$ first neighbors could ensure that the spin-polarized electrical conduction is not interrupted by ASD; (ii) a mixed valence character of the TM ions is typically found on the most prominent HM DPs \cite{Serrate}. Furthermore, it is observed for this compound a magnetic phase separation, with a magnetic transition temperature of $\sim$200 K associated to the FIM coupling between Mn and Ru, and an AFM-type transition at $\sim$90 K attributed to Mn$^{4+}$--O--Mn$^{4+}$ interaction brought by ASD \cite{Ricciardo,Jorgensen}. Although FIM behavior is still clearly observed experimentally, the concomitant presence of such a magnetic inhomogeneity is surely detrimental for long-range order and HM conductivity, resulting in a $T_C$ fairly below that found for archetypical HM DPs.

Interestingly, the substitution of Ca by Sr leads to a more degenerate Mn$^{3+}$ + Ru$^{5+}$ $\leftrightarrow$ Mn$^{4+}$ + Ru$^{4+}$ valence that results in a nearly Mn$^{3.5+}$/Ru$^{4.5+}$ mixed valences state in Sr$_2$MnRuO$_6$ \cite{Ricciardo,King}. However, the incorporation of a relatively larger Sr$^{2+}$ ionic radius into A-site leads to the stabilization of a tetragonal crystal structure, in which orbital ordering along the elongated $c$ axis gives rise to a cooperative Jahn-Teller distortion that results in a C-type AFM ordering below $\sim$200 K and non-metallic electrical conductivity \cite{Ricciardo,Moodenbaugh,Kolesnik}. At the same time, electronic structure calculations indicate the presence of both majority and minority spin states at the Fermi level, making this a non-HM material \cite{Mishra,Moodenbaugh}.

In order to overcome the aforementioned drawbacks of the A$_2$MnRuO$_6$ DPs, we have developed here the Ca$_{1.5}$La$_{0.5}$MnRuO$_6$ (CLMRO) compound, for which the electron doping brought by the partial substitution of La$^{3+}$ for Ca$^{2+}$ in Ca$_2$MnRuO$_6$ can be anticipated to deliver the desired mixed valences on the TM ions. At the same time, the similar ionic radii of Ca$^{2+}$ and La$^{3+}$ preserve the octahedral $a^{-}b^{+}a^{-}$ rotations of the orthorhombic $Pnma$ structure that prevents the C-type AFM ordering. Furthermore, unlike the other HM DPs such as Sr$_2$Fe(Mo,Re)O$_6$, where the electron doping-induced ASD is detrimental for the electrical conductivity and spin polarization, this hindrance may not occur with the La$^{3+}$-doping in the already disordered Ca$_2$MnRuO$_6$. This is because FM-favored coupling and good conductivity are expected for both Ru--Ru \cite{Zhou,Kolesnik,Cao} and mixed-valent Mn--Mn interactions \cite{Jorgensen}.

Here, the structural, electronic and magnetic properties of a polycrystalline sample of CLMRO are investigated by means of x-ray powder diffraction, high resolution transmission electron microscopy (HRTEM), x-ray absorption and magnetic circular dichroism at the Mn-$L_{2,3}$ and Ru-$M_{2,3}$ edges, DC magnetometry, electrical transport and first principles calculations based on density functional theory (DFT). Our results reveal a near room temperature FIM, HM material exhibiting intergrain tunneling magnetoresistance comparable to that of the most known HM DPs. These characteristics are discussed in terms of the mechanisms of orbital hybridization between the TM ions.

\section{Experimental details}

The polycrystalline CLMRO sample was synthesized by conventional solid-state reaction. Stoichiometric amounts of CaO, La$_{2}$O$_{3}$, MnO and RuO$_2$ were mixed and heated at $1100^{\circ}$C for 36 hours in air atmosphere. Then, the material was grinded and sent to $1350^{\circ}$C for 48 hours, with intermediate grinding and pelletizing. After this procedure, a dark black disk of $\sim$10 mm diameter $\times$ $\sim$2 mm thickness was obtained. The crystal structure was verified by means of high-resolution XRD experiment carried at room temperature using a Bruker D8 Discover diffractometer,
operating with  Cu $K_{\alpha}$ radiation. The XRD data was taken over the angular range $10^{\circ}\leq2\theta\leq90^{\circ}$, with a 2${\theta}$ step size of 0.01$^{\circ}$. Rietveld refinement was performed using GSAS software and its graphical interface program \cite{GSAS}.

The morphology, particle sizes and crystallinity of the sample were analyzed using a JEOL Transmission Electron Microscope (TEM) (model JEM 2100) working at 200 kV. The powder of the sample was dispersed in 3 mL of n-propanol and sonicated for 5 min. Droplets of this dispersion were placed over a copper grid coated with parlodion and carbon films and dried in air.

The magnetic measurements were carried out using a Quantum Design's Physical Property Measurement System (PPMS). The temperature dependent magnetization [$M(T)$] curves were carried out in both zero-field-cooled (ZFC) and field-cooled (FC) modes, while the magnetization as a function of magnetic field [$M(H)$] measurements were carried out after ZFC the sample. The electrical transport measurements as a function of temperature and applied magnetic field were also carried out in the PPMS, in a nearly cuboid shaped pellet of $\sim$5$\times$2$\times$1 mm dimensions. The four-contact configuration was adopted, using platinum wires and silver paste for the contacts. 

The Mn $L_{2,3}$- and Ru $M_{2,3}$-edges XAS and XMCD measurements were carried out at beamline I06 at Diamond Light Source. For these experiments, powdered fractions of the CLMRO pellet were spaded on conductive carbon tapes and measured in total electron yield (TEY). For XMCD, the sample was cooled at 2 K, and an external magnetic field of $\pm$ 6 T was applied along the beam axis. 

\section{Computational details}

We used the projector augmented wave (PAW) method along with the spin-polarized generalized gradient approximation (GGA) to DFT as implemented in the Vienna Ab-initio Simulation Package (VASP) \cite{Kresse,Kresse2,Hohenberg,Kohn,Perdew}. In our calculations, a plane-wave cutoff energy of 500 eV was employed. We used the PAW-GGA pseudopotentials (La, Ca$_{sv}$, Mn, Ru, and O) from the VASP distribution with valence configurations of 5s$^2$5p$^6$5d$^1$6s$^2$ for La, 3$s^2$3$p^6$4$s^2$ for Ca, 3$d^6$4$s^1$ for Mn, 4$d^7$5$s^1$ for Ru, and 2$s^2$2$p^4$ for O \cite{Kresse3}.  To sample the k-space Brillouin zone, a 6$\times$6$\times$6 mesh in the Monkhorst-Pack scheme was used \cite{Monkhorst}. All atoms in the system were fully relaxed until the residual forces on the atoms were smaller than 0.01 eV/\AA. 

To study the properties of CLMRO in the presence of complete chemical disorder, we use a special quasi-random structure (SQS) \cite{Wei}, constructed with the \textit{gensqs} program of the alloy theoretic automated toolkit \cite{Walle}. Such scheme has been successfully used to simulate random disorder within small periodic supercells convenient for DFT studies \cite{Padilha,Rashid}. The supercells used contained 80 atoms.

Also, strongly correlated transition-metal oxides have localized $d$-electrons which have energies near the Fermi level (E$_F$). It is well known that local density approximation (LDA) or generalized gradient approximation (GGA) within DFT framework often fail to describe the electronic and magnetic properties correctly. A popular and efficient way to treat the static correlations is the DFT + Hubbard U method, introduced by Anisimov et al. \cite{Anisimov}, which has allowed the study of many TM materials with significant improvements over LDA and GGA. To improve the speed of our calculations and thus facilitate the use of large supercells, we used the rotationally invariant Dudarev approach to GGA+U \cite{Dudarev}, in which only one effective Hubbard parameter U$_{eff}$ = U - J is used, with U and J being Hubbard repulsion and intra-atomic exchange, respectively, for the electrons in the localized $d$ states. The U$_{eff}$ were 5.50 and 3.00 for Mn and Ru, respectively. Similar values of U$_{eff}$ were previously employed \cite{Saad}.

\section{Results and discussion}

\subsection{x-ray diffraction and microscopy}

\begin{figure}
\begin{center}
\includegraphics[width=0.49 \textwidth]{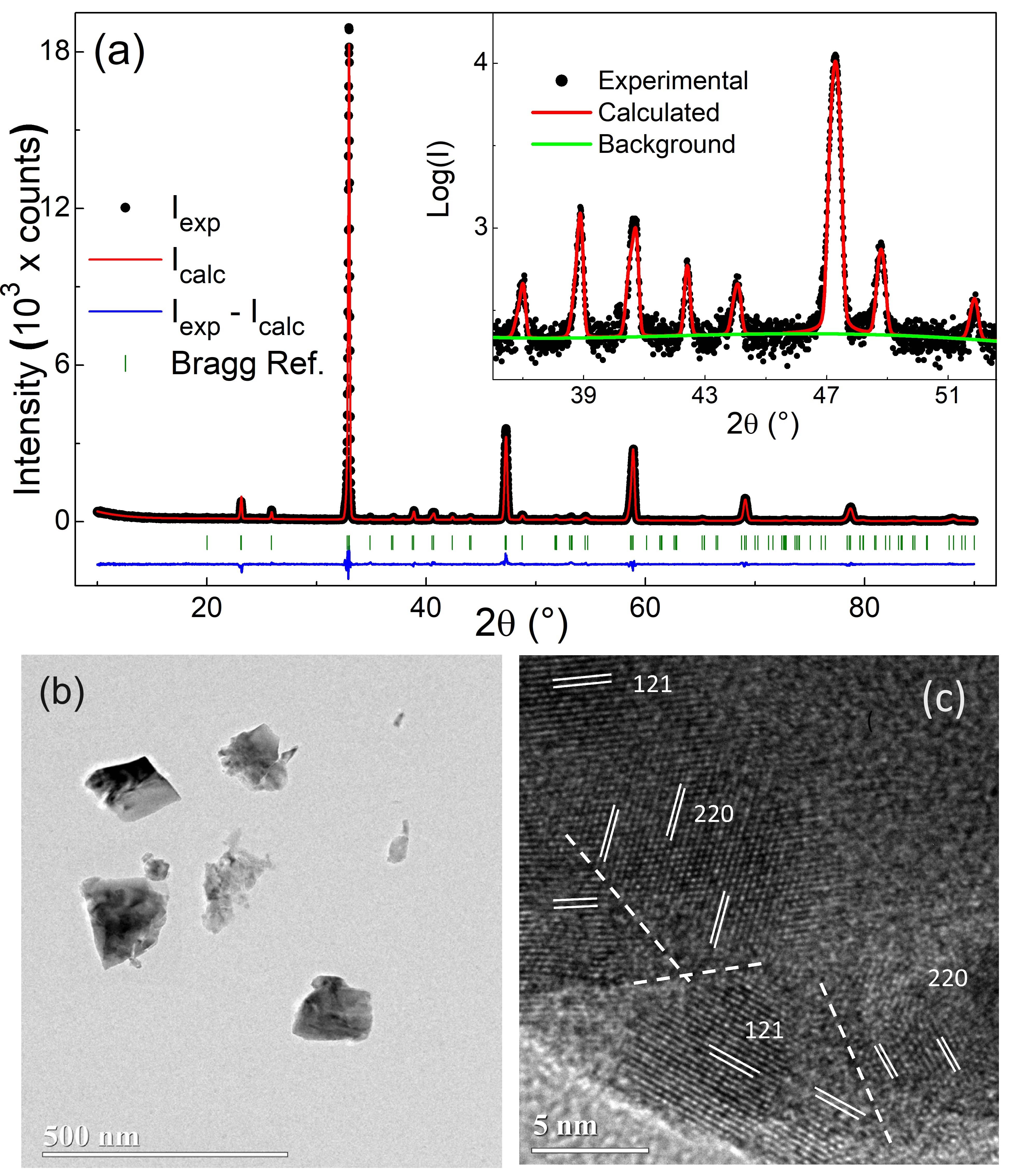}
\end{center}
\caption{(a) Rietveld refinement fitting of the room temperature XRD within the $Pnma$ space group. The inset shows a magnified view of a stretch of the observed and calculated intensities in a Log scale, highlighting the reliability of the refinement. (b) TEM image at 20000$\times$ magnification of the CLMRO powder. (c) HRTEM image of a grain at 600000$\times$ magnification. The solid lines represent Bragg planes identified in the figure and the numbers are their respective Miller indices, while the dashed lines refer to intergrain regions.}
\label{Fig_XRD}
\end{figure}

Fig. \ref{Fig_XRD}(a) shows the Rietveld refinement fitting of the room temperature XRD pattern of CLMRO. No trace of superstructure Bragg reflections corresponding to ordered arrangement of Mn and Ru was found. The observed and calculated intensities depicted in log-scale at the inset of Fig. \ref{Fig_XRD}(a) indicate a well-match with a single phase perovskite belonging to the $Pnma$ space group, as endorsed by the weighted profile $R$-factor $R_{wp}$ = [$\sum_i w_i(I_{exp,i}-I_{calc,i})^2/\sum_i w_{i}I_{exp,i}^2]^{1/2}$ and the profile $R$-factor $R_p$ = $\sum_i |I_{exp,i}-I_{calc,i}|/\sum_i I_{exp,i}$ displayed in Table \ref{T1}. For this  orthorhombic space group, Mn and Ru share the same crystallographic site, meaning that these ions are randomically distributed along the oxygen octahedra, in contrast to the archetypal ordered A$_2$FeMoO$_6$ and A$_2$FeReO$_6$ HM perovskites \cite{Serrate}. On the other hand, this is in agreement with previous reports for similar MnRu-based perovskites such as Ca$_2$MnRuO$_6$ and CaLaMnRuO$_6$ \cite{Ricciardo,King,Granado}, being consistent with the fact that the valence and ionic radii mismatch between Mn$^{3+/4+}$ and Ru$^{4+}$ are much smaller than that between Fe$^{2+/3+}$ and Mo$^{5+/6+}$ on the FeMo-based and between Fe$^{2+/3+}$ and Re$^{5+/6+}$ on the FeRe-based compounds \cite{Serrate,Shannon}. These parameters are known to control the ASD in DPs \cite{Sami,Serrate,Alam}. Attempts to refine the occupancies at the oxygen sites did not improve the quality of the fitting. Although care must be taken in such analysis due to the weak oxygen scattering factor for Cu-$K_{\alpha}$ radiation, the results are consistent with previous neutron powder diffraction (NPD) and thermogravimetric studies on resemblant MnRu-based perovskites, which revealed minimal to no deviations from full occupancies at the O-sites \cite{King,Moodenbaugh,Granado,Ramesha}. As a result, it can be inferred that our CLMRO sample has nearly full O-stoichiometry.

The lattice parameters, depicted in Table \ref{T1}, lies in between those reported for Ca$_2$MnRuO$_6$ \cite{Ricciardo,King} and CaLaMnRuO$_6$ \cite{King,Granado}, as expected due to the slightly larger ionic radius of La$^{3+}$ with respect to Ca$^{2+}$ (1.36 and 1.34 \AA, respectively \cite{Shannon}). Contrasting with the Jahn-Teller-distorted structure of Sr$_2$MnRuO$_6$, for which the axial and equatorial Mn/Ru--O bond lengths are remarkably different from each other \cite{Ricciardo}, the bond distances found for our CLMRO sample (1.9776(1), 1.9718(1) and 1.9828(1) \AA) are fairly close to each other. This results is similar to that of Ca$_2$MnRuO$_6$, for which there was no indication of Jahn-Teller effect \cite{Ricciardo,Mishra}. In contrast, the Mn--O--Ru bond angles of CLMRO, 152.27(1)$^{\circ}$ and 153.23(1)$^{\circ}$, are slightly smaller than those reported for Ca$_2$MnRuO$_6$ \cite{King,Lufaso}, which may impact the Mn--O--Ru orbital hybridization.

\begin{table}
\caption{Main results obtained from the Rietveld refinement with $Pnma$ space group. The numbers in parenthesis correspond to the standard deviations in the last digits.}
\label{T1}
\begin{tabular}{ccccc}
\hline \hline
$a$ (\AA) & $b$ (\AA) & $c$ (\AA) & $R_{wp}$ (\%) & $R_p$ (\%) \\
5.4553(1) & 7.6799(2) & 5.4259(1) & 11.3 & 7.6 \\
\hline

Atom & Wickoff position & $x$ & $y$ & $z$\\

Ca/La & 4$c$ & 0.0262(2) & 0.25 & 0.0039(5)  \\

Mn/Ru & 4$b$ & 0 & 0 & 0.5  \\

O$_1$ & 4$c$ & -0.0225(5) & 0.25 & 0.5891(8)  \\

O$_2$ & 8$d$ & 0.2122(6) & -0.0444(4) & 0.7871(9)  \\

\hline \hline
\end{tabular}
\end{table}

Fig. \ref{Fig_XRD}(b) shows a TEM image of CLMRO. As can be seen, the grains present about 100-200 nm size and uneven shape resulting from the grinding of the pellet. A closer inspection of each grain, achieved by HRTEM images such as that depicted in Fig. \ref{Fig_XRD}(c), reveals that each large grain is formed by several smaller and coalesced grains of approximately 10 nm diameter. The solid lines in the figure represent some identified Bragg planes, while the dashed lines refer to intergrain boundary zones in which the change in direction of the crystal planes between neighboring grains is obvious. Since the boundary region between grains is characterized by natural crystalline disorder \cite{Ernst}, this localized disorder consequently affects the transport and magnetic properties of CLRMO, as will be further discussed at subsequent sections.

\subsection{XAS and XMCD measurements}

\begin{figure}
\begin{center}
\includegraphics[width=0.48 \textwidth]{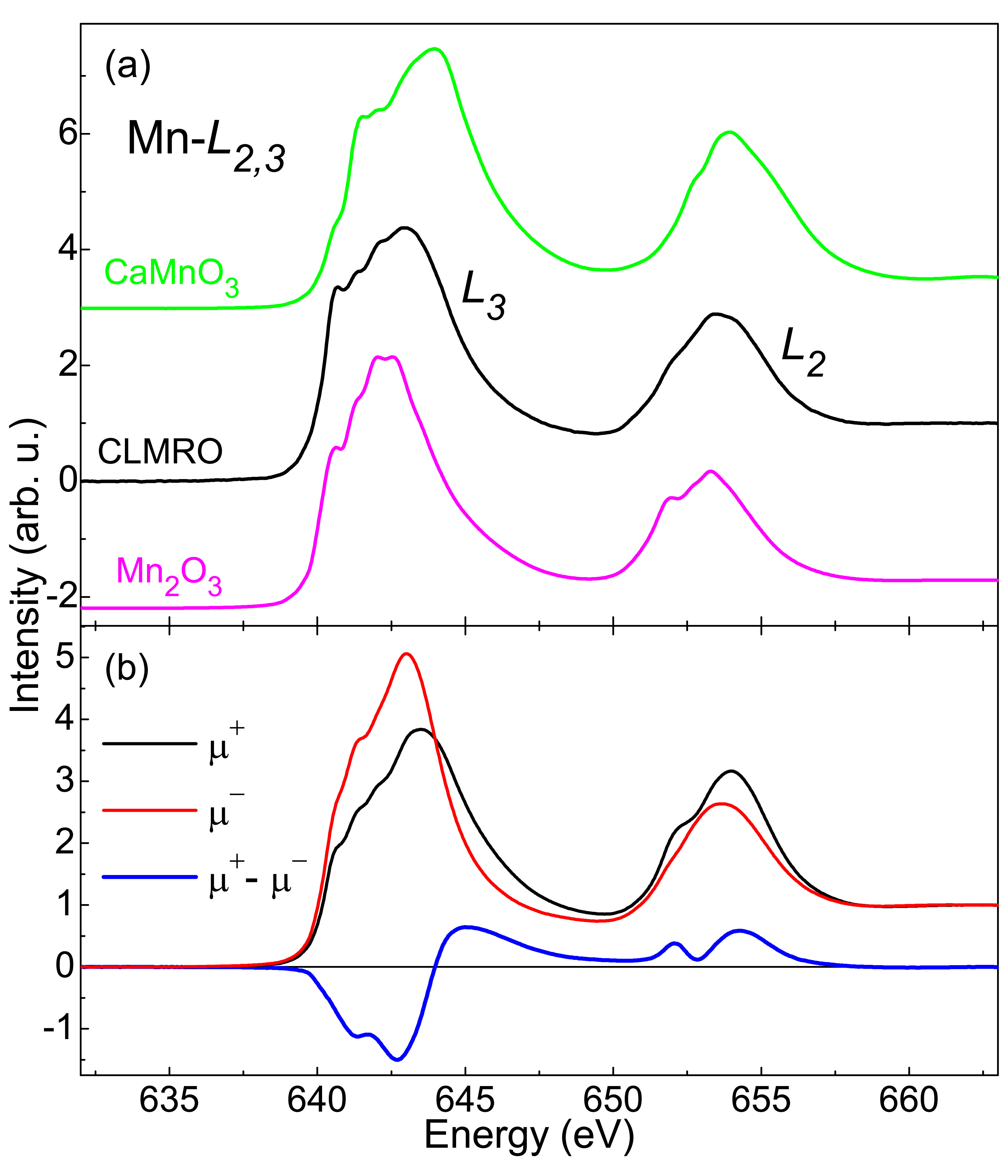}
\end{center}
\caption{(a) Normalized room temperature Mn $L_{2,3}$-edge XAS of CLMRO, together with the spectra of Mn$_2$O$_3$ and CaMnO$_3$, references for Mn$^{3+}$ and Mn$^{4+}$, respectively. (b) Mn-$L_{2,3}$ spectra of CLMRO taken with circularly polarized x-rays at 2 K. $\mu^+$ and $\mu^-$ represent the absorption cross section for photon spin aligned parallel and antiparallel to the applied magnetic field, respectively.}
\label{Fig_XAS_Mn}
\end{figure}

The normalized Mn $L_{2,3}$-edge XAS spectra, related to the 2$p^{6}$3$d^{n}$ to 2$p^{5}$3$d^{n+1}$ absorption process, can bring valuable insights about the Mn electronic configuration in our system. As depicted in Fig. \ref{Fig_XAS_Mn}(a), the spectral features of CLMRO are not the same as the individual line shapes of Mn$_2$O$_3$ and CaMnO$_3$, used here as references for Mn$^{3+}$ and Mn$^{4+}$ valence states, respectively. Indeed, it is equivalent to a mixture of the spectra of the reference samples. Furthermore, it is important to notice that on moving from Mn$_2$O$_3$ to CaMnO$_3$, the absorption spectra shifts toward higher energies, with the CLMRO spectra lying in between those of the references for Mn$^{3+}$ and Mn$^{4+}$ valences. These features sign to a Mn$^{3+}$/Mn$^{4+}$ mixed valence state in our sample. 

In order to verify the magnetic structure of CLMRO, we carried out Mn-$L_{2,3}$ XMCD at 2 K, with an applied magnetic field $H$ = 6 T [Fig. \ref{Fig_XAS_Mn}(b)]. The resulting XMCD signal is fairly similar to those previously reported for Mn in octahedral symmetry, with the $L_{3}$ and $L_{2}$ signals exhibiting opposite signs \cite{Koide,Kuepper,Terai,Burnus,Coutrim,Freeland}. In particular, for Mn$^{3+}$ it is expected a nearly single $L_{3}$-edge signal with a subtle shoulder at the low energy side of the curve \cite{Koide}, while for mixed-valent Mn$^{3+/4+}$ this shoulder starts to get more pronounced \cite{Koide,Kuepper,Terai}, and finally for Mn$^{4+}$ two well separated signals are observed \cite{Burnus,Coutrim,Freeland}. In the case of CLMRO, the line shape confirms the Mn$^{3+/4+}$ mixed valence. 

The sum rules analysis of $L_{2,3}$ XMCD spectra are often employed to obtain quantitative estimates of the orbital ($m_{l}$) and spin ($m_{s}$) magnetic moments of TM ions \cite{Thole,Carra,Chen}. This analysis can be summarized in the following equations
\begin{equation}
m_{l}=- \frac{4\int_{L_3+L_2}(\mu^+-\mu^-)d\omega}{3\int_{L_3+L_2}(\mu^++\mu^-)d\omega}N_{h}, \label{Eq9}
\end{equation} 
\begin{equation}
\begin{split}
m_{s} & = -\frac{6\int_{L_3}(\mu^+-\mu^-)d\omega-4\int_{L_3+L_2}(\mu^+-\mu^-)d\omega}{\int_{L_3+L_2}(\mu_++\mu_-)d\omega}\times \\
& N_{h}\left(1+ \frac{7\langle T_z\rangle}{2\langle S_z \rangle}\right)^{-1},  \label{Eq10}\\
\end{split}
\end{equation} 
where $\mu^{+}$ and $\mu^{-}$ represent respectively the absorption cross section for photon spin aligned parallel and antiparallel to $H$, $S_z$ denotes the projection along $z$ of the spin magnetic momentum, $N_h$ represent the number of empty 3$d$ states, $T_z$ denotes the magnetic dipole moment, and $L_2$ and $L_3$ represent the integration ranges. 

Neglecting $T_z$ and using $N_h$ = 6.5, which is an approximated atomistic value corresponding to an equal mixture of Mn$^{3+}$ ($N_h$ = 6) and Mn$^{4+}$ ($N_h$ = 7), we obtain $m_{l}$ = 0.006 $\mu_B$/atom and  $m_{s}$ = 1.472 $\mu_B$/atom. The small $m_{l}$ value is in agreement with a nearly quenched orbital contribution for Mn. On the other hand, the $m_{s}$ value here found is significantly smaller than one could expect for a $\sim$Mn$^{3.5+}$ configuration. Part of this discrepancy may be ascribed to imprecision in the determination of $N_h$. For instance, if single Mn$^{4+}$ valence configuration is assumed, $m_{s}$ and $m_{l}$ increase to 1.582 and 0.007 $\mu_B$/atom, respectively, while for single Mn$^{3+}$ one gets $m_{s}$ = 1.359 $\mu_B$/atom and $m_{l}$ = 0.006 $\mu_B$/atom. Additionally, several studies have questioned the validity of sum rules calculations for Mn due to the spectral overlap at the $L_{2,3}$ edges \cite{Teramura,Groot,Goering}. Furthermore, for CLMRO, a significant electron delocalization is expected, as it will be discussed in the context of the electrical transport data. It is important to note that the $T_z$ = 0 assumption may not be a reasonable approximation for Mn \cite{Groot}. 

In any case, such moments are comparable to previous sum rule-derived values reported for Mn in octahedral symmetry \cite{Koide,Kuepper,Garcia,Pal}. In fact, the total magnetic moment found, $m_{t}$ = 1.48 $\mu_B$/atom, is smaller than that reported for Mn$^{3+}$ in LaMnO$_3$ \cite{Kuepper,Garcia}, as well as that reported for Mn$^{3.4+}$ and Mn$^{3.5+}$ in La$_{1-x}$Sr$_x$MnO$_3$ \cite{Koide,Kuepper}, however it is larger than that of Mn$^{4+}$ in La$_2$NiMnO$_6$ \cite{Pal}. Since the Mn magnetic moment is expected to decrease as its oxidation state increases from +3 to +4, these results further suggest a mixed valence in our CLMRO sample, with a slight majority of tetravalent Mn ions.

\begin{figure}
\begin{center}
\includegraphics[width=0.48 \textwidth]{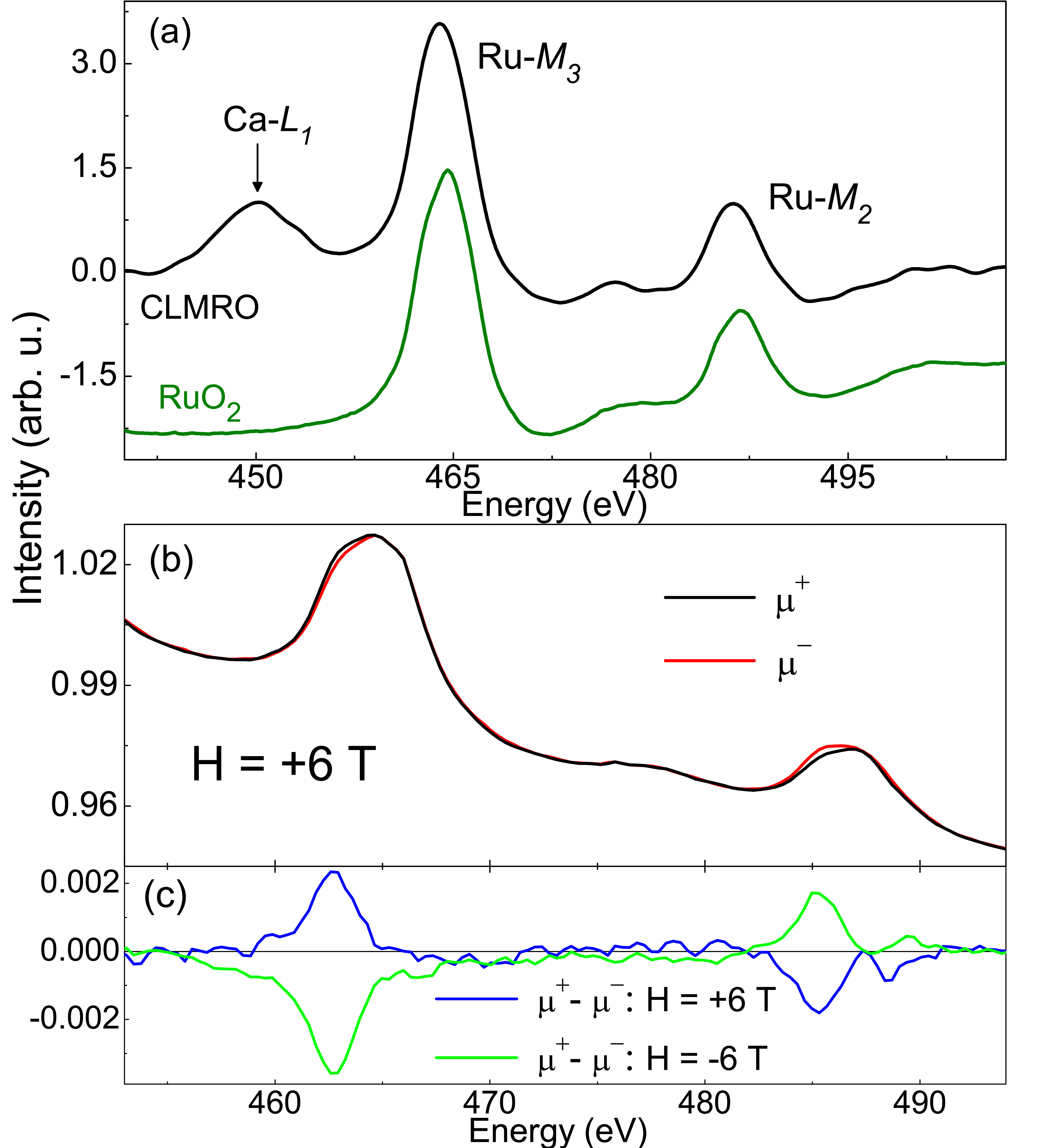}
\end{center}
\caption{(a) Room temperature Ru-$M_{2,3}$ XAS spectra for CLMRO. The spectra of RuO$_2$ is also depicted as reference for Ru$^{4+}$ configuration. (b) Ru $M_{2,3}$-edge spectra of CLMRO taken with circularly polarized x-rays at 2 K. The photon spin was aligned parallel ($\mu^+$, black solid) and antiparallel ($\mu^-$, red dashed) to the +6 T magnetic field. (c) Ru $M_{2,3}$-edge XMCD ($\mu^+$ - $\mu^-$) for CLMRO, carried out at 2 K with $H$ = +6 T (blue) and -6 T (green).}
\label{Fig_XAS_Ru}
\end{figure}

The Ru $M_{2,3}$-edge XAS, associated to transitions from the 3$p_{3/2}$ and 3$p_{1/2}$ core levels into the Ru 4$d$ band, are displayed in Fig. \ref{Fig_XAS_Ru}(a). The peak at $\sim$450 eV corresponds to the Ca $L_1$-edge absorption. The line shape of the Ru $M_{2,3}$ XAS is broad due to the more extended character of the 4$d$ orbitals as well as the relatively short core-hole lifetime of the 3$p$ level \cite{Terai}, making it challenging to estimate the valence of Ru. Nevertheless, by comparing the XAS spectrum of CLMRO with that of RuO$_2$, used as reference for the Ru$^{4+}$ state, we note a significant similarity between the spectral features. This indicates that the majority of Ru ions present tetravalent state in CLMRO. Indeed, for a stoichiometric Ca$_{1.5}$La$_{0.5}$MnRuO$_6$ compound, the nearly equivalent proportions of Mn$^{3+}$ and Mn$^{4+}$ would lead to a configuration that is nearly Ru$^{4+}$. 

The Ru $M_{2}$ and $M_{3}$ dichroic signals of the XMCD data obtained with $H$ = +6 T [blue line in Fig. \ref{Fig_XAS_Ru}(b)], are opposite to those of Mn $L_{2}$ and $L_{3}$. This suggests AFM coupling between Mn and Ru that would result in FIM behavior on CLMRO, as observed for Ca$_2$MnRuO$_6$ \cite{Ricciardo,Mishra}. In spite of the small Ru-XMCD signal observed, the spectrum carried out with $H$ = -6 T [green line in Fig. \ref{Fig_XAS_Ru}(b)] shows a reversal in sign with respect to that measured with $H$ = +6 T, demonstrating that it represents an authentic response of the material to an external field. The small Ru $M_{2,3}$ XMCD observed could be ascribed to strong orbital hybridization with the neighboring O ions, causing a significant electron delocalization. The possible presence of some few Ru$^{3+}$ ions could also contribute to the decrease of the dichroic signal. However, such a small XMCD is most likely related to competing magnetic couplings induced by the ASD. Our DFT calculations suggest that Ru are AFM coupled with the Mn ions, whereas FM coupled with their nearest neighbor Ru ions. 

Interestingly, the peak position of the $M_3$-edge XMCD (462.7 eV) is shifted toward lower energy with respect to that of the $M_3$-edge XAS (464.1 eV). Similar behavior found in SrRuO$_3$ films was ascribed to the fact that, while the XAS peak comes from the transitions to both the Ru 4$d$ $t_{2g}$ and $e_g$ bands, the XMCD signal is mainly associated with the $t_{2g}$ states \cite{Agrestini}. Y. K. Wakabayashi \textit{et al.} further interpreted this same result in SrRuO$_3$ films as evidence of spin polarization of the $t_{2g}$ states at the Fermi level \cite{Yuki,Takiguchi}, which is endorsed by ab initio calculations performed for this compound as well as for other Ru-based perovskites \cite{Mishra,Rondinelli}. In this sense, these results can be viewed as a first indicative of HM behavior in CLMRO. 

The sum rules analysis for Ru $M_{2,3}$-edges is particularly challenging due to the smaller absorption cross section at these edges as a result of the lower transition probabilities compared to $L_{2,3}$-edges. This leads to a low signal-to-noise ratio for XMCD at $M_{2,3}$-edges, significantly impacting the calculation of the orbital and spin moments. An additional challenge for CLMRO is the fact that the $M_{2,3}$-edges overlap with the Ca $L_1$-edge, making the baseline subtraction and integration of XAS data unreliable. For these reasons we have not shown the sum rules analysis for Ru $M_{2,3}$-edges. 

\subsection{Magnetometry}

\begin{figure}
\begin{center}
\includegraphics[width=0.47 \textwidth]{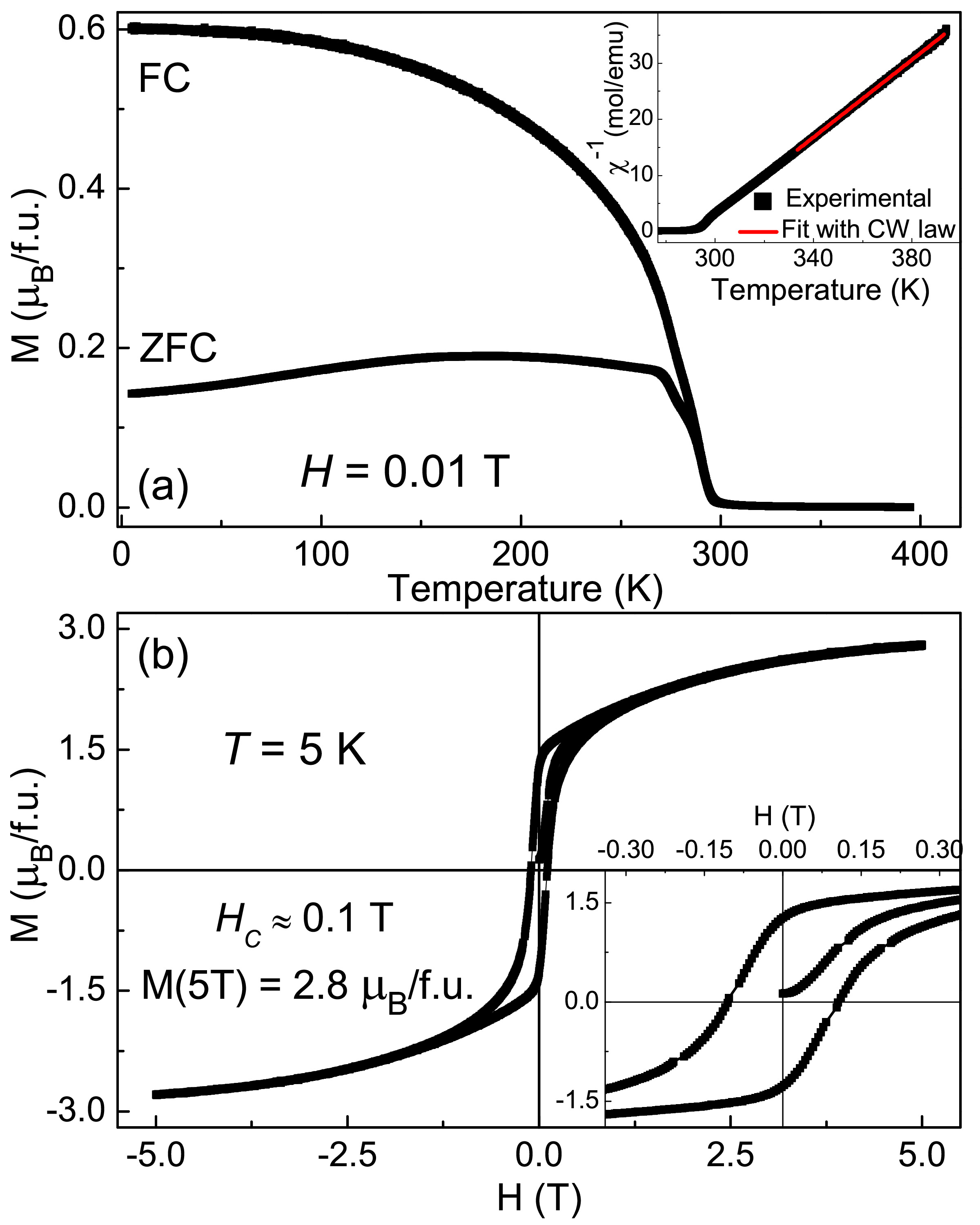}
\end{center}
\caption{(a) ZFC-FC $M(T)$ curves measured with $H$ = 0.01 T. The inset shows the ZFC $\chi^{-1}$ curve, where the straight line corresponds to best fit of PM region with the CW law. (b) $M(H)$ curve measured at 5 K, where the inset shows a magnified view of the low $H$ region.}
\label{Fig_PPMS}
\end{figure}

The ZFC-FC $M(T)$ curves of CLMRO, displayed in Fig. \ref{Fig_PPMS}(a), exhibit FM- or FIM-like shape. Nonetheless, the magnitude of the magnetization at low temperature is much smaller than the value expected for FM configuration, pointing to the FIM picture suggested by the XMCD results. As will be discussed in section F, this is most likely associated to the Mn($e_g$)--O(2$p$)--Ru($t_{2g}$) orbital hybridization. The ZFC curve exhibits a subtle increase of the magnetization with decreasing temperature below the magnetic transition, which may be related to the presence of ASD-induced Mn--O--Mn and Ru--O--Ru interactions. It may also be contributed by the FM-favored Mn($t_{2g}$)--O(2$p$)--Ru($t_{2g}$) hybridization, which is expected to become more relevant as the Ru--O--Mn bond angle increases at higher temperatures.

The magnetic ordering temperature, defined as the inflection point of $M(T)$, is $T_C$ $\simeq$ 291 K. This represents a remarkable increase with respect to the $\sim$200 K transition temperatures observed for Ca$_2$MnRuO$_6$ and CaLaMnRuO$_6$ \cite{Ricciardo,Granado,Taniguchi}. For other DPs such as the widely investigated HM Sr$_{2-x}$La$_x$FeMoO$_6$ \cite{Navarro} and Sr$_{2-x}$La$_x$FeReO$_6$ \cite{Blasco} series, the electron doping caused by partial La$^{3+}$ for Sr$^{2+}$ substitution also leads to large increases in $T_C$ (of about 70 K/$e^{-}$ for the FeMo- and 140 K/$e^{-}$ for the FeRe-based system), ascribed to the strengthening of the orbital hybridization between the TM ions due to the band  filling at the Fermi level \cite{Serrate}. However, for these systems, as well for the great majority of ordered DPs, the electron doping is unavoidably accompanied by an increase in ASD, which suppress the spin polarization and decreases the magnetization since the nearest neighbor 3$d$--3$d$ interaction is usually AFM. 

In the case of the Ca$_{2-x}$La$_x$MnRuO$_6$ series, the remarkable increase in the $T_C$, of $\sim180$ K/$e^{-}$ from $x$ = 0 to 0.5, followed by a decrease of approximately the same amount from $x$ = 0.5 to 1.0, can be understood in terms of both the electron hopping-driven mechanism and the evolution of the Mn valence state. For Ca$_2$MnRuO$_6$, Mn is mainly tetravalent \cite{Ricciardo}, while in CaLaMnRuO$_6$ it is trivalent \cite{Mishra}. In both cases, the Mn--Mn interactions are expected to be AFM. In contrast, for CLMRO ($x$ = 0.5), the XAS and XMCD results show a Mn$^{3+}$/Mn$^{4+}$ mixed-valent state, and the Mn$^{3+}$--Mn$^{4+}$ interaction is a prototypical double-exchange FM coupling. In such case, the Mn$^{3+}$--O--Mn$^{4+}$ path is not detrimental for the electron hopping or to the long-range order, thus the magnetic coupling is strengthened via kinetic energy-driven mechanism.

The inverse of magnetic susceptibility ($\chi^{-1}$) as a function of temperature is displayed at the inset of Fig. \ref{Fig_PPMS}(a). The fit of the paramagnetic region with the Curie-Weiss (CW) law yields a CW temperature $\theta_{CW}$ = 290 K, matching with $T_C$. Its positive signal, however, suggests a predominance of FM interactions on the material. Despite the fact that our XMCD, DFT and low temperature $M(T)$ data indicate antiparallel orientation of nearest neighbor Mn and Ru moments, we anticipate that the DFT calculations also suggest that the Mn--O--Mn and Ru--O--Ru interactions brought by ASD are of FM nature. Furthermore, as aforementioned the FM-favored Mn($t_{2g}$)--O(2$p$)--Ru($t_{2g}$) hybridization may play a part at higher temperatures. Similar situation was already found for other HM DPs, such as the archetypal Sr$_2$FeMoO$_6$, on which the CW fit suggests FM interaction \cite{Martinez}, whereas further investigation by means of NPD, XMCD and nuclear magnetic resonance demonstrated FIM behavior resulting from AFM coupling between the 3$d$ and 5$d$ ions \cite{Sanchez,Besse,Kapusta}.

The effective magnetic moment obtained from the CW fit is $\mu_{eff}$ = 4.8 $\mu_B$/f.u.. From this, we can speculate about the TM ion valences by adapting the usual equation for systems presenting two or more magnetic ions, $\mu$ = $\sqrt{{\mu_1}^2 + {\mu_2}^2 + ...}$ \cite{PRB2020,JPCM2021}, to the following
\begin{equation}
\begin{split}
\mu & = [(0.5-x){\mu_{Mn^{3+}}}^2 + (0.5+x){\mu_{Mn^{4+}}}^2 \\
 &  + (1-x){\mu_{Ru^{4+}}}^2 + x{\mu_{Ru^{3+}}}^2]^{1/2}. \label{Eq_mu} \\
\end{split}
\end{equation}

For the Mn$^{3+}$ and Mn$^{4+}$ moments, we can assume the spin-only approximation-derived values (4.9 and 3.9 $\mu_B$, respectively \cite{Ashcroft}), that are indeed very close to the values reported for LaMnO$_3$ (5.0 $\mu_B$ \cite{Balbashov}) and CaMnO$_3$ (3.9 $\mu_B$ \cite{Trokiner}). For Ru, however, the spin-only may not be a good approximation due to possible non-negligible orbital contribution and spin-orbit coupling. Therefore, we used for Ru$^{4+}$ and Ru$^{3+}$ the moments reported for CaRuO$_3$ ($\sim2.5$ $\mu_B$ \cite{Felner,Singh}) and LaRuO$_3$ ($\sim2.0$ $\mu_B$ \cite{Sinclair,Bouchard}), from which Eq. \ref{Eq_mu} yields the presence of $\sim$ 25\% of Ru$^{3+}$ in our sample. Surely, this is just a rough estimate, since deviations from the individual moments assumed are expected to occur.

As for the $M(T)$ curves, the $M(H)$ loop depicted in Fig. \ref{Fig_PPMS}(b) is also typical of FM- or FIM-like systems, with a coercive field of $\sim$0.1 T and a remanent magnetization $M_{R}$ $\simeq$ 1.3 $\mu_B$/f.u.. Despite the fact that the curve shows a tendency to saturation, some slope persists at the high field region, and the "saturated'' magnetization, \textit{i.e.} the magnetization value at the maximum applied field, is of approximately 2.8 $\mu_B$/f.u., which is much smaller than the value expected for a FM system composed of Mn and Ru. These findings provide further evidence of FIM behavior resulting from AFM coupling between Mn and Ru. The ground state magnetization value expected for AFM coupling between Mn$^{3+}$ (S = 4/2)/Mn$^{4+}$ (S = 3/2) and Ru$^{3+}$ (S = 1)/Ru$^{4+}$ (S = 1/2) can be estimated by the following equation
\begin{equation}
M = g[(0.25\times\frac{4}{2} + 0.75\frac{3}{2})-(0.75\times\frac{2}{2} + 0.25\times\frac{1}{2})]\simeq1.5 \mu_B/f.u., \ \label{Eq_Ms}
\end{equation}
where the TM ion's fractions assumed are those obtained from the CW-fit, and the Land\'{e} factor is considered as $g\simeq2$. The calculated value is reasonably close to $M_R$, and the discrepancy between the remanent and the "saturated'' magnetizations can be ascribed to the presence of AFM couplings in a system presenting a highly disordered distribution of the TM ions along the lattice. As our DFT calculations will show (section E), although the nearest neighbor Mn--Mn and Ru--Ru couplings brought by ASD are most likely FM, the resulting moment of these pairs can be initially coupled antiparallel, and it is possible to expect a progressive flip of these moments toward field direction with increasing $H$. As mentioned earlier, the FM character of these ASD-driven couplings is a unique aspect of this material compared to typical HM DPs, and will impact its transport properties, as will be discussed next.

\subsection{Magnetotransport}

\begin{figure}
\begin{center}
\includegraphics[width=0.50 \textwidth]{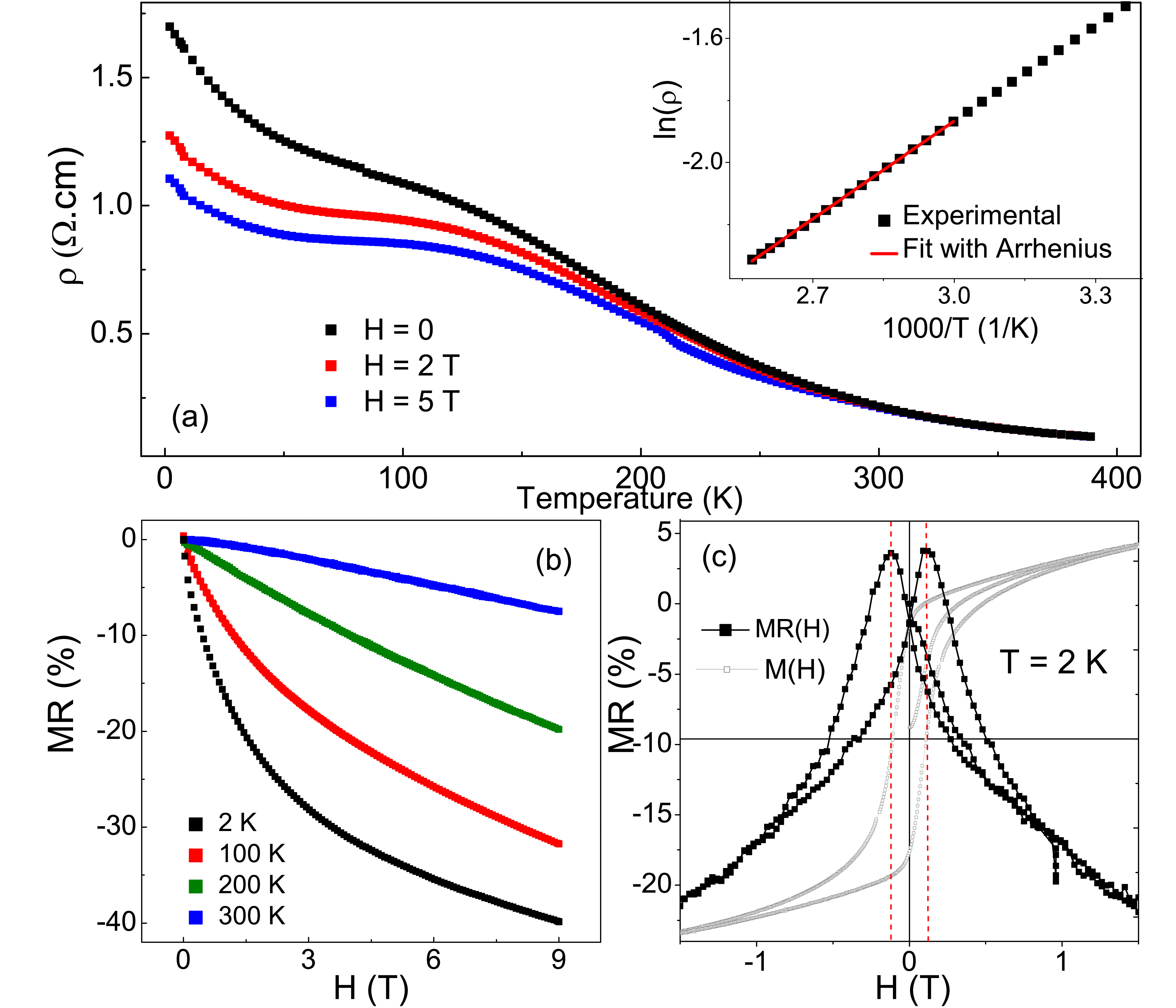}
\end{center}
\caption{(a) $\rho(T)$ measured with $H$ = 0, 2 T and 5 T. The inset shows the fit of the high temperature region of the $H$ = 0 curve with the thermally activated model. (b) $MR(H)$ curves measured at some selected temperatures. (c) Magnified view of the low $H$ region of a field-cycled $MR(H)$ curve measured at 2 K, where a $M(H)$ curve is also reproduced for comparison. The vertical lines are guides for the eye.}
\label{Fig_rho}
\end{figure}

The electrical resistivity ($\rho$) and magnetoresistance ($MR$) measurements performed on CLMRO are displayed in Fig. \ref{Fig_rho}. Qualitatively, the overall increase of $\rho$ with decreasing temperature suggests a semiconducting-like behavior, with $\rho \simeq$ 0.2 $\Omega$cm at room temperature. However, the small variation of $\rho$ over the whole temperature-interval contrasts with the usual logarithmic evolution observed on semiconducting and insulating DPs \cite{Serrate,Sami,Moodenbaugh}. A fit of the high temperature region with the power law equation of the thermally activated mechanism, inset of Fig. \ref{Fig_rho}(a), yields an energy gap $E_{G}\simeq0.1$ eV which is smaller than usually found for semiconductors. Furthermore, it is noticed a change in the slope of the $\rho(T)$ curves, with a plateau-like region followed by a more pronounced increase of $\rho$ at lower temperatures. This is typical of grain boundary effects in polycrystals, where the highly defective nature of the grain's surface can lead to insulating behavior even when the core is metallic, with the spin-dependent scattering usually becoming relevant at temperatures much below $T_C$ due to the highly frustrated character of the surface \cite{Serrate,Ritter}. From this, we speculate that CLMRO might be intrinsically metallic, the gentle increase of $\rho$ with decreasing temperature resulting from grain boundary effects.

As Fig. \ref{Fig_rho}(a) shows, $\rho(T)$ curves measured with distinct magnetic fields start to deviate from each other below $T_C$, demonstrating a close connection between the magnetic and transport properties in CLMRO. Fig. \ref{Fig_rho}(b) depicts isothermal $MR(H)$ measurements carried out at some selected temperatures. The 2 K curve shows steepest slope at the low-field region, followed by a roughly linear dependence at higher fields. Again, this is a characteristic feature of HM polycrystalline DPs exhibiting negative intergrain tunnelling $MR$. In this case, the linear behavior is caused by the HM bulk, while in the low field region, grain boundary effects dominate. When a field is applied, it can align the initially frustrated spins, consequently reducing the spin-dependent scattering \cite{Serrate,Prado}. The $MR \simeq$ -40\% observed at $H$ = 9 T on the 2 K curve is comparable to that of archetypal HM DPs such as Sr$_2$FeMoO$_6$ ($MR \sim$ -30\% at 4.2 K with $H$ = 7 T) \cite{Tokura} and Sr$_2$FeReO$_6$ ($MR \sim$ -25\% at 10 K with $H$ = 6 T) \cite{Marest}.

A magnified view of the low field region of a field-cycled $MR(H)$ curve measured at 2 K is depicted in Fig. \ref{Fig_rho}(c), where its butterfly-like shape gives further evidence that intergrain tunnelling $MR$ plays a major role on the magnetotransport properties of CLMRO \cite{Tokura,Serrate,Marest}. For comparison, a $M(H)$ curve measured at 2 K is also displayed in Fig. \ref{Fig_rho}(c). As the vertical lines show, the maxima in the $MR$ reasonably match the coercive field of the $M(H)$ curve, indicating that the slope of the $MR$ curve changes its signal only when the internal spins are aligned with the external field. 

\subsection{Electronic Structure Calculations}

\begin{figure*}
\begin{center}
\includegraphics[width= \textwidth]{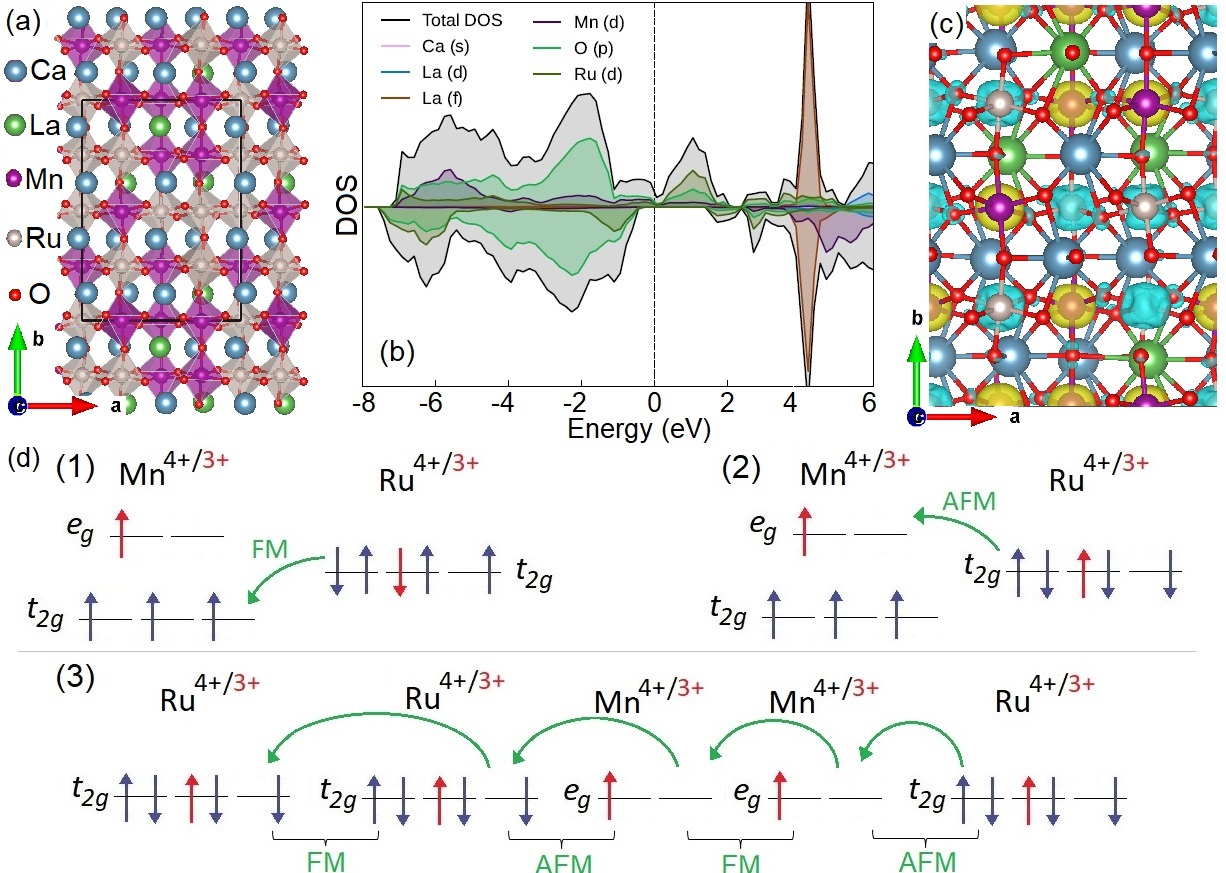}
\end{center}
\caption{(a) Eight f.u. supercell of disordered CLMRO, where the blue spheres represent the Ca ions, the green ones are for the La, the purple are for Mn, the grey for Ru, and the red ones for O. The black lines indicate the unitcell. (b) Total and orbital-resolved projected density of states for the disordered (i) configuration (see text). (c) Spin densities distribution for the (i) configuration of disordered CLMRO. The spheres follow the same color-code of panel (a), while the light yellow (blue) contour corresponds to the majority (minority) spin density. The isosurface value is 0.04 (-0.04) e/\AA$^{-3}$. (d) Schematic diagram of the mechanisms of hybridization between Mn$^{4+}$/Mn$^{3+}$ and Ru$^{4+}$/Ru$^{3+}$. The intervening O 2$p$ orbitals were omitted, for simplicity, and the red arrows represents additional spins for Mn$^{3+}$ and Ru$^{3+}$, which are absent in Mn$^{4+}$ and Ru$^{4+}$. Panel (1) represents a FM $t_{2g}$--$t_{2g}$ interaction, panel (2) is for an AFM $t_{2g}$--$e_g$ coupling, and (3) shows the spin-dependent electrical transport along the CLMRO lattice within the (i) disordered configuration}.
\label{Fig_DFT}
\end{figure*}

To study the structural, electronic and magnetic properties of CLMRO in the presence of complete chemical disorder, we use an 8 f.u. SQS structure as shown in Fig. \ref{Fig_DFT}(a), where four Mn (Mn$_{\textrm{Mn}}$) and four Ru (Ru$_{\textrm{Ru}}$) atoms occupy their correct lattice sites whereas the remaining eight octahedral sites are occupied by an equal number of Mn$_{\textrm{Ru}}$ (Mn atoms occupying the Ru sublattice) and Ru$_{\textrm{Mn}}$ (Ru atoms occupying the Mn sublattice) antisites, a protocol similar to the previously employed by Mishra \textit{et al.} for obtaining a disordered Ca$_{2}$MnRuO$_{6}$ \cite{Mishra}. 

The DFT-PBE calculated lattice parameters were: $a$ = 5.495 \AA, $b$ = 7.645 \AA, $c$ = 5.368 \AA, with a super-cell volume of 225.51 \AA$^3$ in close agreement with the experimental result of 226.42 \AA$^3$. Also, the disorder introduces some variation in the lengths of the Mn--O bonds (ranging from 1.889 to 2.109 \AA) and Ru--O bonds (ranging from 1.887 to 2.069 \AA), with calculated average bonds of 2.009 and 1.969 \AA, respectively. The calculated average angle Mn/Ru-O-Mn/Ru was 150.47$^{\circ}$, which is only 2.42$\%$ smaller than the experimental result of 154.11$^{\circ}$.

To understand the coupling between the magnetic moments of the Mn and Ru ions in the presence of disorder, we studied the energetic stability of five different conformations. All the investigated conformations resulted in spin-polarized density of states at the Fermi level. Table \ref{T2} shows the energy ($E$) per f.u. of the considerd magnetic configurations along with their magnetization ($M$). Our results clearly indicate the configuration (i) to be energetically most stable, followed by configuration (ii), which is 37 meV/f.u. higher in energy. In addition to the energetic stability, configuration (i) shows a magnetic moment of 1.56 $\mu_B$/f.u., reasonably close to the experimentally measured $M_R\simeq$ 1.3 $\mu_B$/f.u..

\begin{table}
\caption{Magnetic conformations, Energy per formula unit ($E$) and Magnetization ($M$) for SQS-disordered CLMRO.}
\label{T2}
\begin{ruledtabular}
\begin{tabular}{c|ccc|c|c}
Conform. & \multicolumn{3}{c|}{Magnetic interactions} & $E$ (eV/f.u.) & $M$ ($\mu_B$/f.u.) \\
\hline  
 & Mn--Ru & Mn--Mn & Ru--Ru &  &   \\           
 (i) &  AFM   & FM    & FM  & 0.000   &  1.56 \\
 (ii) &  AFM  & FM    & AFM  &  0.037  &  2.76  \\
 (iii) & AFM  & FM    & AFM   &  0.063 & 3.94  \\ 
 (iv) & AFM  & AFM & FM   &  0.135 &  2.16  \\
 (v) &  FM  & FM    & FM  &  0.145 &  6.34  
\end{tabular}
\end{ruledtabular}
\end{table}

For the most energetically stable structure, configuration (i), we are going to discuss in more detail its electronic structure. In Fig. \ref{Fig_DFT}(b) the total density of states (DOS) and the orbital-resolved density of states (PDOS) are presented. We find a significant density of Ru-$d$ and Mn-$d$ states crossing the Fermi level in the spin up channel, whereas the spin down channel presents a bandgap of $\approx$ 1.66 eV. This makes disordered CLMRO a 100 $\%$ spin-polarized HM system. Interestingly, there are also O-$p$ states crossing the Fermi level in the spin up channel, which is important for opening the paths for the electron hopping.

Contour plots (isosurfaces) of the spin density are depicted in Fig. \ref{Fig_DFT}(c) for the magnetic configuration (i). The Figure clearly shows that the magnetic moments arise mainly from Mn and Ru atoms. In addition, the calculated magnetic moments for Mn and Ru are 3.5 and 1.8 $\mu_B$/ion, respectively. Thus, approximately, these moments correspond to Mn$^{3.5+}$ and Ru$^{3.8+}$ oxidation states. The La and Ca atoms are essentially nonmagnetic and the spin density in the interstitial region is negligible. This is in agreement with the previously discussed results on the calculated magnetic moments. Besides, one can see small spin density accumulated on O atoms, giving further indication of strong $p$-$d$ hybridization between O and the TM ions.

\subsection{Discussion}

From the set of experimental and theoretical results described above, we are now able to discuss the mechanisms of hybridization responsible for the magnetotransport properties of CLMRO. In double-perovskites, the orbital hybridization between the $d$ states, mediated by O 2$p$ orbitals, is responsible for the effective (or virtual, in the case of insulating behavior via superexchange interaction) spin-dependent electron hopping between one TM ion to another, which can stabilize either FM or AFM coupling \cite{Sami,Tokura,Kobayashi,Serrate,Dasgupta,Tang,PRB2020,Philipp}. In octahedral symmetry, Mn assumes high spin configuration, with Mn$^{4+}$ presenting its $t_{2g}$ orbitals partially filled and its $e_g$ orbitals empty, while for Mn$^{3+}$ there is one additional electron at the $e_g$ level. Conversely, the stronger crystal field splitting in Ru stabilizes it in the low spin configuration. Thus, for Ru$^{4+}$ (Ru$^{3+}$) there is one (two) completely filled and two (one) partially filled $t_{2g}$ orbital(s), while both $e_g$ orbitals are empty.

According to previous studies, there are two primary exchange paths between Mn and Ru, namely the Mn($t_{2g}$)--O(2$p$)--Ru($t_{2g}$) and the Mn($e_g$)--O(2$p$)--Ru($t_{2g}$) hybridizations \cite{Mishra,Ricciardo,Terai}. As can be seen in Fig. \ref{Fig_DFT}(d), the spin-dependent electron hopping through the first path (panel 1) stabilizes a FM-favored coupling between Mn and Ru, while panel 2 secures an AFM-favored coupling. Since our experimental and theoretical results indicate FIM behavior in CLMRO, we conclude that the Mn($e_g$)--O(2$p$)--Ru($t_{2g}$) hybridization channel is the most relevant for our sample. Here it is important to mention that, although this coupling is in principle forbidden by symmetry in a cubic perovskite (\textit{i.e.} for a 180$^{\circ}$  Mn--O--Ru bond angle), it becomes possible due to the octahedral tilts caused by the orthorhombic lattice distortion. In this sense, CLMRO resembles 3$d$-5$d$ DPs for which the extended character of the 5$d$ orbitals, as well as their inaccessible $e_g$ states, makes 3$d$($e_g$)--O(2$p$)--5$d$($t_{2g}$) the prevalent path of orbital hybridization \cite{Morrow,PRB2023,Feng}.

The bottom panel of Fig. \ref{Fig_DFT}(d) schematizes the electronic hopping in the system within the disordered (i) configuration, \textit{i.e.} considering an AFM coupling between Mn and Ru that results in FIM, while the Mn--Mn and Ru--Ru nearest neighbor interactions brought by ASD are FM. It is clear that such arrangement is not detrimental for spin-dependent electronic transport, evidencing that CLMRO should be a convenient material for fully spin-polarized transport since ordering does not need to be enforced during growth.

In terms of technological applications, the persistence of a robust magnetization above room temperature is desirable for a HM material, meaning that the $T_C$ $\simeq$ 291 K observed for CLMRO is a drawback. On the other hand, the spin-polarized electron hopping between Mn and Ru being not interrupted by the FM-favored Mn--Mn and Ru--Ru interactions brought by ASD, together with the significant increase of $T_C$ with partial La for Ca substitution, demonstrate a way to tune the magnetotransport properties of this class of HM materials. This can boost the investigation of similar MnRu-based DPs in order to achieve even higher $T_C$.

\section{Summary}

Here, we describe the synthesis and a thorough investigation of the structural, electronic and magnetic properties of Ca$_{1.5}$La$_{0.5}$MnRuO$_6$. Our results show that this is a fully disordered perovskite, formed in orthorhombic $Pnma$ space group, presenting a $T_C$ = 291 K associated to the Mn--O--Ru AFM coupling, that results in FIM behavior. Its remerkably higher ordering temperature with respect to that of Ca$_{2}$MnRuO$_6$ ($T_C\simeq$ 200 K) can be ascribed to the La$^{3+}$ doping-induced mixed valence character of the TM ions, on which the nearest neighbor Mn--O--Mn and Ru--O--Ru FM interactions are not detrimental for the kinetic energy-driven long-range magnetic order. Our electric resistivity measurements combined with DFT calculations evidence HM transport, with a low temperature negative magnetoresistance comparable to that of archetypal double-perovskites, which is ascribed to grain boundary effects. The main Mn($e_g$)--O--Ru($t_{2g}$) AFM orbital hybridization, together with the ASD-driven Mn($e_g$)--O-- Mn($e_g$) and Ru($t_{2g}$)--O--Ru($t_{2g}$) FM ones, are believed to open the path for spin-dependent transport in Ca$_{1.5}$La$_{0.5}$MnRuO$_6$. The combination of near room temperature and spin-polarized conductivity in fully disordered CLMRO highlights the potential of MnRu-based perovskites for applications in spintronics.

\begin{acknowledgements}
This work was supported by the Brazilian funding agencies: Funda\c{c}\~{a}o Carlos Chagas Filho de Amparo \`{a} Pesquisa do Estado do Rio de Janeiro (FAPERJ) [Nos. E-26/202.798/2019 and E-26/211.291/2021], Funda\c{c}\~{a}o de Amparo \`{a}  Pesquisa do Estado de Goi\'{a}s (FAPEG) and Conselho Nacional de Desenvlovimento Cient\'{\i}fico e Tecnol\'{o}gico (CNPq) [Nos. 400633/2016-7, 309599/2021-0 and 305394/2023-1].  R. B. P. acknowledges LaMCAD-UFG, supercomputer SDumont/LNCC-MCTI and Cluster Euler/CeMEAI for providing the computational resources. We thank Diamond Light Source for time on beamline I06 under proposal MM31735-1. 
\end{acknowledgements}

\end{document}